\documentclass[preprint,showpacs,preprintnumbers,amsmath,amssymb]{revtex4}

\usepackage{graphicx}
\usepackage{dcolumn}
\usepackage{bm}

\begin{document}


\title{Neutrino phenomenology of a high scale supersymmetry model}

\author{Ying-Ke Lei} \email{leiyingke@itp.ac.cn} \quad
\author{Chun Liu} \email{liuc@itp.ac.cn}

\address{\small CAS Key Lab. of Theor. Phys., 
Institute of Theoretical Physics,
Chinese Academy of Sciences, Beijing 100190, China \\
and \\
School of Physical Sciences, 
Univ. of Chinese Academy of Sciences, Beijing 100049, China}

\date{\today}

\begin{abstract}
CP violation in the lepton sector, and other aspects of neutrino 
physics, are studied within a high scale supersymmetry model.  In 
addition to the sneutrino vacuum expectation values (VEVs), the heavy 
vector-like triplet also contributes to neutrino masses.  Phases of the 
VEVs of relevant fields, complex couplings and Zino mass are considered.  
The approximate degeneracy of neutrino masses $m_{\nu_1}$ and 
$m_{\nu_2}$ can be naturally understood.  The neutrino masses are then 
normal ordered, $\sim$ 0.020 eV, 0.022 eV, and 0.054 eV.  Large CP 
violation in neutrino oscillations is favored.  The effective Majorana 
mass of the electron neutrino is about 0.02 eV.  
\end{abstract}



\maketitle

\section{Introduction}

Neutrino physics, like leptonic CP violation, is an interesting topic 
\cite{neutrino} in the current research of particle physics.  Among 
other things, it might be the final place where experiments of particle 
physics will give definite results in the near future.  The results will 
check various theoretical models about the fermion masses of the 
Standard Model (SM).

We proposed that supersymmetry (SUSY) \cite{susy} can be the theory 
underlying the fermion masses in Refs. \cite{1,2,3}.  The basic idea is 
the following.  It assumes a flavor symmetry.  The flavor symmetry is 
broken after the sneutrinos obtain nonvanishing vacuum expectation 
values (VEVs).  (In this way, SUSY is motivated.)  These VEVs result in 
a nonvanishing neutrino mass.  The empirical smallness of neutrino 
masses needs very large SM super partner masses to be understood which 
are about $10^{12}$ GeV.  Thus, our SUSY is of high scale breaking 
\cite{meta,hsusy,splitsusy}.  

A further natural assumption is that the flavor symmetry breaks softly.
Namely the soft SUSY breaking masses of the sfermions do not obey the 
flavor symmetry either.  The theoretical reason is that the soft masses 
are due to the supergravity effect which generically breaks any global
symmetry.  Soft breaking of the flavor symmetry implies that the lepton
number violation due to sneutrino VEVs is explicit instead of being
spontaneous.  Therefore there is no any massless Nambu-Goldstone boson
related to nonvanishing sneutrino VEVs.  Actually the large masses of
the model make the low energy effective theory just the SM via Higgs
mass fine tuning, except for that we have an understanding of the
hierarchical pattern of the charged lepton masses, or that of the SM
Yukawa coupling constants.

To briefly review the model in a simple way, the SM is SUSY generalized.
The flavor symmetry is $Z_3$ cyclic among the three generation SU(2)$_L$
lepton doublets $L_1$, $L_2$ and $L_3$.  The other fields are trivial 
under $Z_3$.  The $Z_3$ invariant combinations are $\sum_{i=1}^{3}L_i$ 
and 
$\epsilon_{\alpha\beta}(L_1^{\alpha}L_2^{\beta}+L_2^{\alpha}L_3^{\beta}+L_3^{\alpha}L_1^{\beta})$ 
with $\alpha$ and $\beta$ denoting the SU(2)$_L$ indices.  
In terms of the following redefined lepton superfields,
$L_e     = \displaystyle \frac{1}{\sqrt 2}(L_1-L_2)$, 
$L_\mu   = \displaystyle \frac{1}{\sqrt 6}(L_1+L_2-2L_3)$, 
$L_\tau  = \displaystyle \frac{1}{\sqrt 3}(\sum_i L_i)$,
the above $Z_3$ invariant combinations are $L_\tau$ and 
$\epsilon_{\alpha\beta}L_e^{\alpha}L_\mu^{\beta}$, respectively.  The 
superpotential is then
\begin{equation}
\label{1}
{\mathcal W} \supset y_\tau\epsilon_{\alpha\beta}
L_\tau^{\alpha}H_d^{\beta}E_\tau^c
+\epsilon_{\alpha\beta}L_e^{\alpha}L_\mu^{\beta}
(\lambda_\tau E_\tau^c + \lambda_\mu E_\mu^c)
+\bar{\mu}\epsilon_{\alpha\beta}H_u^{\alpha}H_d^{\beta} \,,
\end{equation}
where $H_u$ and $H_d$ are the two Higgs doublets, the right-handed
lepton singlet $E_\tau^c$ is defined as the one which couples to
$L_\tau$, and $E_\mu^c$ is that orthogonal to $E_\tau^c$ and with a
coupling to $L_eL_\mu$. $y_\tau$, $\lambda_\tau$ and $\lambda_\mu$ are
coupling constants.  (Note that considering the mixing between $L_\tau$
and $H_d$ gives the same form of the above superpotential \cite{2}.)  It
is seen that the electron is massless, because $E_e^c$ is always absent
in the Lagrangian.  This is true whenever SUSY is conserved, the 
nonvanishing electron mass is due to SUSY breaking (together with 
electroweak gauge symmetry and flavor symmetry breaking via loops).  
Note that all the coupling constants in our superpotential are assumed 
to be natural values, say typically $\sim 0.01-1$, and the mass 
parameter $\bar{\mu}$ is taken to be large $\sim 10^{12}$ GeV.  The SM 
fermion mass hierarchy is due to symmetries and their breaking.  

In addition, a heavy vector-like SU(2)$_L$ triplet field $T(\bar{T})$ 
with hypercharge $2(-2)$ needs to be introduced so as to make the Higgs 
mass realistic \cite{3,meta}.  This triplet field also contributes to 
neutrino masses.  In terms of the redefined fields, the flavor symmetric 
superpotential relevant to the triplet $T$ and $\bar{T}$ fields is
\begin{equation}
\label{2}
\begin{array}{lll}
{\mathcal W} & \supset & y^\nu \{L_\tau H_d\} T
      + \lambda^\nu_1 \{L_e L_e + L_\mu L_\mu\} T
      + \lambda^\nu_2 \{L_\tau L_\tau\} T \\
  & & + \lambda^\nu_3 \{H_d H_d\} T
      + \lambda^\nu_4 \{H_u H_u\} \bar{T}
      + M_T T \bar{T} \,
\end{array}
\end{equation}
with $M_T$ the mass $\sim 10^{13}$ GeV.  The braces denote that the two
doublets form an SU(2)$_L$ triplet representation.

The soft SUSY breaking terms in the Lagrangian are in general form which
also break the flavor symmetry \cite{1,2,3}.  All the mass parameters of
the model are taken to be about $10^{12}-10^{13}$ GeV.  The spontaneous 
gauge symmetry breaking of the SM occurs.  Through fine tuning, the 
right electroweak vacuum is obtained.  By including contribution due to 
the triplet field, this model can give reasonable neutrino spectrum and 
the mixing pattern, and predicted the right order of $\theta_{13}$ 
\cite{2,3}.  (The quark sector was considered in Ref. \cite{2}.)  

Roughly speaking about the electroweak symmetry breaking.  There are 
five scalar doublets, the mass parameters are all large $\sim 10^{12}$ 
GeV.  Eigenvalues of their mass-squared matrix are generically large.  
However, one of these values can be exceptional, because it is a 
difference between two large parameters.  It is this difference that 
makes the fine-tuning possible.  Whence the difference is tuned to be 
about $-$(100 GeV)$^2$, correct electroweak symmetry breaking occurs.  
The corresponding eigenstate field is one superposition of the five 
doublets.  It is the only light scalar doublet, and is just the SM 
Higgs field from the point of view of the low energy effective field 
theory.  The SM Higgs gets a VEV is equivalent to that the original two 
Higgses and sleptons get their VEVs \cite{2,3}.

\section{Complex couplings and sneutrino VEVs}

In this paper, we will carefully consider CP violation of the lepton 
sector, and completely analyze the neutrino masses and mixing.  In 
general, the coupling constants are complex, however, because of the 
flavor symmetry, many of them can be made real via field phase 
rotation.  In the superpotential Eq. (\ref{1}) for charged leptons, all 
the couplings can be adjusted to be real.  On the other hand, in the 
superpotential Eq. (\ref{2}) for neutrino masses, the couplings cannot 
be all taken real, as can be seen in the following way.  The mass 
parameters $\bar{\mu}$ and $M_T$ are taken real, thus $H_u$ and $H_d$ 
always have opposite phases, and so do $T$ and $\bar{T}$.  
$\lambda^\nu_2$ is real via rotating the phase of $L_\tau$, 
$\lambda^\nu_4$ is real via rotating $H_u$ (or $\bar{T}$), $y_\tau$ is 
real via $E_\tau^c$, $\lambda_\tau$ real via $L_eL_\mu$ rotating, and 
$\lambda_\mu$ real via $E_\mu^c$.  In such a phase convention, only 
$y^\nu$, $\lambda_1^\nu$ and $\lambda_3^\nu$ can be complex.  The 
$\lambda_1^\nu$ term will contribute to the neutrino masses, which was 
omitted in our previous analysis \cite{3}.    

In the soft SUSY breaking terms, the mass parameters and coupling 
constants are generally complex, and there is no enough freedom to 
rotate all of the phases away.  

The scalar potential relevant to the electroweak symmetry breaking is 
\begin{equation}
\begin{split} 
V=&(|\bar{\mu}|^2+m_{h_u}^2)|h_u|^2+(|\bar{\mu}|^2+m_{h_d}^2)|h_d|^2
+\frac{g^2+g'^2}{8}
(|h_u|^2-|h_d|^2-{\tilde{l}_\alpha}^\dag{\tilde{l}_\alpha})^2\\
&+\frac{g^2}{4}[2|h_u^\dag h_d|^2
+2(h_u^\dag\tilde{l}_\alpha)(\tilde{l}_\alpha^\dag h_u) 
+2(h_d^\dag\tilde{l}_\alpha)(\tilde{l}_\alpha^\dag h_d) \\ 
&-2|h_d|^2({\tilde{l}_\alpha}^\dag {\tilde{l}_\alpha}) 
+(\tilde{l}_\alpha^\dag\tilde{l}_\beta)(\tilde{l}_\beta^\dag\tilde{l}_\alpha) 
-(\tilde{l}_\alpha^\dag\tilde{l}_\alpha)(\tilde{l}_\beta^\dag\tilde{l}_\beta)]
 \\ 
&+(\frac{1}{2}m_{d\alpha}^2h_d^\dag\tilde{l}_\alpha 
+\frac{1}{2}m_{\alpha\beta}^2\tilde{l}_\alpha^\dag\tilde{l}_\beta 
+B_\mu h_u h_d+B_{\mu\alpha}h_u \tilde{l}_\alpha+{\rm h.c.})
\end{split}
\end{equation}
where $g$ and $g'$ are SM gauge coupling constants.  $h_u$ and $h_d$ 
denote the scalar components of $H_u$ and $H_d$, respectively, and 
$\tilde{l}_\alpha$'s left-handed sleptons.  $m_{h_{(u,d)}}^2$, 
$m_{d\alpha}^2$, $m_{\alpha\beta}^2$ and $B_\mu$, $B_{\mu\alpha}$ are 
soft squared masses. 

In considering CP violation of the scalar potential, the essential point 
lies in the soft bilinear terms where the mass parameters are complex.  
Field redefinition of $h_d$ and $\tilde{l}_\alpha$ may remove phases of 
$B_\mu$ and $B_{\mu\alpha}$ respectively, however, the phases of 
$m_{d\alpha}^2$ and off-diagonal terms of $m_{\alpha\beta}^2$ are still 
there.  This means that after the electroweak symmetry breaking, Higgs 
and sneutrino VEVs are complex in general.  (Previously we took all the 
VEVs real.)  In the analysis, we still have the freedom to choose the 
VEV of Higgs field $h_u$ to be real, and VEVs of the Higgs and the 
sneutrino fields are denoted as 
($v_u$, $v_d e^{i\delta_{v_d}}$, $v_{l_e}e^{i\delta_{l_e}}$, 
$v_{l_\mu}e^{i\delta_{l_\mu}}$, $v_{l_\tau}e^{i\delta_{l_\tau}}$) where 
the phases have been explicitly written down.  These VEVs enter the 
lepton mass matrices and thus contribute to CP violation in the leptonic 
mixing.

\section{Neutrino masses}

The sneutrino VEVs result in a nonvanishing neutrino mass, 
\begin{equation}
\label{4}
M^\nu_0 =-\displaystyle\frac{a^2}{M_{\tilde{Z}}e^{i\delta_Z}}
\left(\begin{array}{ccc}
v_{l_e} v_{l_e}e^{2i\delta_{l_e}} &v_{l_e} v_{l_\mu} e^{i(\delta_{l_e}+\delta_{l_\mu})}  &v_{l_e} v_{l_\tau}e^{i(\delta_{l_e}+\delta_{l_\tau})}  \\
v_{l_\mu}v_{l_e}e^{i(\delta_{l_e}+\delta_{l_\mu})} &v_{l_\mu} v_{l_\mu}e^{2i\delta_{l_\mu}} &v_{l_\mu} v_{l_\tau}e^{i(\delta_{l_\mu}+\delta_{l_\tau})} \\
v_{l_\tau} v_{l_e}e^{i(\delta_{l_e}+\delta_{l_\tau})}&v_{l_\mu} v_{l_\tau}e^{i(\delta_{l_\mu}+\delta_{l_\tau})}&v_{l_\tau} v_{l_\tau}e^{2i\delta_\tau}
\end{array}
\right),
\end{equation}
where $a=\sqrt{(g^2+g'^2)/2}$, $M_{\tilde{Z}}$ is the Zino mass which is 
the typical superpartner mass, and the phase of Zino mass term, 
$\delta_Z$, is explicitly written.  This is due to gauge interactions, 
it is natural realization of the type-I seesaw mechanism \cite{type1} 
where the role of right-handed neutrinos is replaced by the Zino. In 
addition, the superpotential (\ref{2}) contributes following neutrino 
masses \cite{3}, 
\begin{equation}
M_1^\nu=-\frac{\lambda_4^\nu v_u^2}{M_T}
\begin{pmatrix}
\lambda_1^\nu e^{\delta_{\lambda_1}} & 0 & 0\\
0 &\lambda_1^\nu e^{\delta_{\lambda_1}} & 0\\
0 & 0 & \lambda_2^\nu
\end{pmatrix}\,,
\end{equation}  
where the phase of coupling $\lambda_{1}^\nu$ has been explicitly 
written.  This part of neutrino mass generation is realization of the 
type-II seesaw mechanism \cite{type2}.  

The full neutrino mass matrix is
\begin{equation} 
M^\nu = M_0^\nu + M_1^\nu \,.
\end{equation} 
Note this is the full neutrino mass matrix of the model.  It is due to 
tree level contribution of lepton number violation.  The loop level 
contribution due to R-parity violation is negligible \cite{2}, because 
the sparticles in the loops are very heavy.

The physics analysis including $\lambda_1^\nu$ is different from our 
previous one \cite{3}.  We observe that it is natural to take that 
$M_1^\nu$ is numerically dominant over $M_0^\nu$, then there appears a 
degeneracy between the first two neutrinos.  This roughly fits the 
neutrino spectrum obtained from neutrino oscillation experiments.  This 
degeneracy is perturbed by $M_0^\nu$ which also contributes neutrino 
mixing.  Furthermore, it is interesting to note that inclusion of 
$\lambda_1^\nu$ in certain cases does not really increase difficulty in 
the analysis because $M_1^\nu$ is diagonal.  

We rewrite $M^\nu$ by adjusting the diagonal part $M_1^\nu$ to be 
proportional to identity matrix, 
\begin{equation} 
M^\nu = \tilde{M}_0^\nu + \tilde{M}_1^\nu \,,
\end{equation}
where 
\begin{equation}
\begin{split}
\tilde{M}_0^\nu= & -\frac{a^2}{M_{\tilde{Z}}e^{i\delta_Z}}
 \begin{pmatrix}
 v_{l_e}v_{l_e}e^{i2\delta_{l_e}}&v_{l_e}v_{l_\mu}e^{i(\delta_{l_e}+\delta_{l_\mu})}&v_{l_e}v_{l_\tau}e^{i(\delta_{l_e}+\delta_{l_\tau})}\\
 v_{l_\mu}v_{l_e}e^{i(\delta_{l_e}+\delta_{l_\mu})}&v_{l_\mu}v_{l_\mu}e^{2i\delta_{l_\mu}} &v_{l_\mu}v_{l_\tau}e^{i(\delta_{l_\tau}+\delta_{l_\mu})}\\
 v_{l_\tau}v_{l_e}e^{i(\delta_{l_e}+\delta_{l_\tau})}&v_{l_\tau}v_{l_\mu}e^{i(\delta_{l_\tau}+\delta_{l_\mu})}&v_{l_\tau}v_{l_\tau}e^{2i\delta_{l_\tau}}+\Delta \lambda e^{i(\delta_{\lambda}+\delta_Z)}
 \end{pmatrix},   
\end{split}
\end{equation}  
and 
\begin{equation}
\begin{split}
\tilde{M}_1^\nu= & -\frac{a^2}{M_{\tilde{Z}}}
 \begin{pmatrix}
 \lambda_1^\prime e^{i\delta_{\lambda_1}} & 0 & 0 \\
 0 & \lambda_1^\prime e^{i\delta_{\lambda_1}} & 0 \\
 0 & 0 & \lambda_1^\prime e^{i\delta_{\lambda_1}}
 \end{pmatrix}\,,
\end{split}
\end{equation}
where 
$\lambda_1^\prime=\displaystyle\frac{M_{\tilde{Z}}}{a^2}\frac{\lambda_1\lambda_4v^2_u}{M_T}$, 
$\lambda_2^\prime=\displaystyle\frac{M_{\tilde{Z}}}{a^2}\frac{\lambda_2\lambda_4 v^2_u}{M_T}$, 
and 
$\Delta\lambda e^{i\delta_\lambda}=\lambda_2^\prime-\lambda^\prime_1 e^{i\delta_{\lambda_1}}$.  
Generally, $M^\nu$ is complex, the phases make further analytical 
calculation \cite{7} difficult.  For illustration and an easy analysis, 
and without losing generality about CP violation, we simply take 
$\delta_{l_\alpha}=0$ and $\delta_\lambda=-\delta_Z$ in the following.  
Then, up to an overall factor, $\tilde{M}^\nu_0$ is a real symmetric 
matrix and can be diagonalized by an orthogonal matrix. It just needs 
diagonalizing $\tilde{M}^\nu_0$, because $\tilde{M}_1^\nu$ is 
essentially an unit matrix which does not affect this diagonalization.  
By further assuming that $v_{l_e}^2+v_{l_\mu}^2 \ll v_{l_\tau}^2$ which 
is reasonable because 
$v_{l_\tau}=\displaystyle\frac{v_1+v_2+v_3}{\sqrt{3}}$ which does not 
violate the $Z_3$ flavor symmetry, it is found that $\tilde{M}^\nu_0$ 
is diagonalized by, 
\begin{equation}
O_\nu\simeq\\
\begin{pmatrix}
\displaystyle\frac{v_{l_\mu}}{\sqrt{v_{l_e}^2+v_{l_\mu}^2}}&
\displaystyle\frac{v_{l_e}}{\sqrt{v_{l_e}^2+v_{l_\mu}^2}}&
\displaystyle\frac{v_{l_e}v_{l_\tau}}{v_{l_\tau}^2+\Delta\lambda}\\
-\displaystyle\frac{v_{l_e}}{\sqrt{v_{l_e}^2+v_{l_\mu}^2}}&
\displaystyle\frac{v_{l_\mu}}{\sqrt{v_{l_e}^2+v_{l_\mu}^2}}&
\displaystyle\frac{v_{l_\mu}v_{l_\tau}}{v_{l_\tau}^2+\Delta\lambda}\\ 
0&-\displaystyle\frac{\sqrt{v_{l_e}^2+v_{l_\mu}^2}v_{l_\tau}}
{v_{l_\tau}^2+\Delta\lambda}&1 
\end{pmatrix}  
\end{equation}  
with eigenvalues 
\begin{equation}
\tilde{M}^{\nu{\rm ~ diag}}_0 \simeq \displaystyle 
-\frac{a^2}{M_{\tilde{Z}}}e^{-i\delta_Z}
\begin{pmatrix}
0&0&0\\
0&\displaystyle(v_{l_e}^2+v_{l_\mu}^2)
 \frac{\Delta\lambda}{v_{l_\tau}^2+\Delta\lambda}&0\\
0&0&\displaystyle v_{l_\tau}^2+\Delta\lambda
\end{pmatrix}\,.  
\end{equation}  
  
In fact, $O_\nu$ diagonalizes $M^\nu$, 
\begin{equation}
{O_\nu}^T M^\nu O_\nu = 
\tilde{M}_0^{\nu ~ {\rm diag}} + \tilde{M}_1^\nu \,.
\end{equation}  
Noticing that the diagonalized matrix is still complex, we further 
write that 
\begin{equation}
\label{Mnu}
\begin{split}
&\tilde{M}_0^{\nu ~ {\rm diag}} +\tilde{M}_1^\nu \\[0.5cm] 
=& -\frac{a^2}{M_{\tilde{Z}}}
\begin{pmatrix}
e^{i\frac{\delta_{\lambda_1}}{2}}&0&0\\
0&e^{i\frac{\beta_1}{2}}&0\\
0&0&e^{i\frac{\beta_2}{2}}
\end{pmatrix}
\begin{pmatrix}
m_{\nu_1}       & 0         & 0 \\
0               & m_{\nu_2} & 0 \\
0               & 0         & m_{\nu_3}
\end{pmatrix}
\begin{pmatrix}
e^{i\frac{\delta_{\lambda_1}}{2}}&0&0\\
0&e^{i\frac{\beta_1}{2}}&0\\
0&0&e^{i\frac{\beta_2}{2}}
\end{pmatrix}, 
\end{split}
\end{equation}
the neutrino masses in our model are 
\begin{equation} 
\label{m3}
\begin{array}{lll}
m_{\nu_1} & = & 
\displaystyle\frac{a^2}{M_{\tilde{Z}}}\lambda^\prime_1 \,,\\[3mm]
m_{\nu_2} & \simeq &\displaystyle\frac{a^2}{M_{\tilde{Z}}}
[\lambda_1^\prime+(v_{l_e}^2+v_{l_\mu}^2)
\displaystyle\frac{\Delta\lambda}{v_{l_\tau}^2+\Delta\lambda}
\cos(\delta_{\lambda_1}+\delta_Z)] \,,\\[3mm] 
m_{\nu_3} & = & \displaystyle\frac{a^2}{M_{\tilde{Z}}}
\sqrt{{\lambda^\prime_2}^2+v_{l_\tau}^4
+2\lambda^\prime_2 v_{l_\tau}^2 \cos\delta_Z} \\, 
\end{array}
\end{equation}  
with the phases 
\begin{equation}
\begin{split}
 &\beta_1 \simeq \delta_{\lambda_1},\\
 &\beta_2=\arctan\frac{v_{l_\tau}^2\sin{\delta_Z}}
{\lambda^\prime_2 + v_{l_\tau}^2\cos{\delta_Z}}\,.  
\end{split}
\end{equation}

It is clear that $\nu_1$ and $\nu_2$ are almost degenerate with a mass 
$\simeq\displaystyle\frac{a^2}{M_{\tilde{Z}}}\lambda_1^\prime$.  Their 
mass splitting is about 
$\displaystyle\frac{a^2}{M_{\tilde{Z}}}(v_{l_e}^2+v_{l_\mu}^2)
\displaystyle\frac{\Delta\lambda}{v_{l_\tau}^2+\Delta\lambda}$.  
$v_{l_\tau}^2+\Delta\lambda$ and $\Delta\lambda$ have the same order of 
magnitude by definition, and we take 
$(v^2_{l_e}+v^2_{l_\mu})\simeq\displaystyle\frac{\lambda^\prime_1}{10}$.  
According to neutrino oscillation experiments \cite{pdg}, 
$\Delta m^2_{12}  = 8.0 \times 10^{-5}$ eV$^2$ and 
$|\Delta m^2_{23}|= 2.4 \times 10^{-3}$ eV$^2$, 
this model typically gives that 
\begin{equation}
m_{\nu_1}\simeq 2.0 \times 10^{-2} {\rm eV}\,, ~~~ 
m_{\nu_2}\simeq 2.2 \times 10^{-2} {\rm eV}\,, ~~~
m_{\nu_3}\simeq 5.4 \times 10^{-2} {\rm eV}\,.
\end{equation}  
Naturally the phases in above formulae are $\mathcal{O}(1)$.  
This makes us to take all the cosines to be $\mathcal{O}(1)$ for 
simplicity in estimating the neutrino masses.  And $m_{\nu_3}$ is 
numerically fixed by choosing $\lambda_2^\prime$ and $v_{l_\tau}^2$.  

Finally, we obtain the unitary matrix $U_\nu$ which diagonalizes 
$M^\nu$, 
\begin{equation}
U_\nu^T M^\nu U_\nu = -\frac{a^2}{M_{\tilde{Z}}}
\begin{pmatrix}
m_{\nu_1} & 0               & 0 \\
0               & m_{\nu_2} & 0 \\
0               & 0         & m_{\nu_3}
\end{pmatrix}, 
\end{equation}  
\begin{equation}
U_\nu=O_\nu P^\dag  
\end{equation}
with $P$ being the pure phase matrix appearing in Eq. (\ref{Mnu}).

\section{Charged lepton masses}

From Eq. (\ref{1}), the charged lepton mass matrix is obtained.  
Considering the sneutrino and Higgs VEVs are complex, it is 
\begin{equation}
M^l=
\begin{pmatrix}
0&\lambda_\mu v_{l_\mu}e^{i\delta_{l_\mu}}&\lambda_\tau v_{l_\mu}e^{i \delta_{l_\mu}}\\
0&\lambda_\mu v_{l_e}e^{i\delta_{l_e}}&\lambda_\tau v_{l_e}e^{i \delta_{l_e}}\\
0&0                                    & y_\tau v_de^{i\delta_{v_d}}
\end{pmatrix}.
\end{equation}  
Here the electron mass is neglected.  In this model, the electron mass 
would be a loop contribution of SUSY breaking terms which also break the 
flavor symmetry and the electroweak symmetry \cite{1,2}.  $M^l$ in the 
above equation basically fixes the mixing due to charged leptons with a 
precision of $m_e/m_\mu$.  It is standard to find the unitary matrix 
$U_l$ which diagonalizes 
$M^l {M^l}^\dag$, 
\begin{equation}
U_l^\dag M^l {M^l}^\dag U_l=
\begin{pmatrix}
m_e^2 & 0               & 0 \\
0          & m_\mu^2 & 0 \\
0          & 0               & m_\tau^2 
\end{pmatrix}.
\end{equation} 
It can be expressed as 
\begin{equation*}
U_l=P_lO_l \,,
\end{equation*}
where 
\begin{equation}
P_l=\begin{pmatrix}
e^{i\delta_{l_\mu}}& 0                       & 0 \\
0                           & e^{i\delta_{l_e}} & 0 \\
0                           & 0                        & e^{i\delta_{v_d}}
\end{pmatrix}, 
\end{equation}  
\begin{equation}
O_l\simeq\begin{pmatrix} 
\displaystyle\frac{-v_{l_e}}{\sqrt{v_{l_e}^2+v_{l_\mu}^2}}&
\displaystyle\frac{v_{l_\mu}}{\sqrt{v_{l_e}^2+v_{l_\mu}^2}}
\displaystyle\frac{y_\tau v_d}{\sqrt{y_\tau^2 v_d^2+\lambda_\tau^2
(v_{l_e}^2+v_{l_\mu}^2)}}&
\displaystyle\frac{\lambda_\tau v_{l_\mu}}{\sqrt{y_\tau^2 v_d^2+
\lambda_\tau^2(v_{l_e}^2+v_{l_\mu}^2)}}\\
\displaystyle\frac{v_{l_\mu}}{\sqrt{v_{l_e}^2+v_{l_\mu}^2}}&
\displaystyle\frac{v_{l_e}}{\sqrt{v_{l_e}^2+v_{l_\mu}^2}}
\displaystyle\frac{y_\tau v_d}{\sqrt{y_\tau^2 v_d^2+\lambda_\tau^2
(v_{l_e}^2+v_{l_\mu}^2)}}&
\displaystyle\frac{\lambda_\tau v_{l_e}}{\sqrt{y_\tau^2 v_d^2+
\lambda_\tau^2(v_{l_e}^2+v_{l_\mu}^2)}}\\
0&\displaystyle\frac{-\lambda_\tau\sqrt{v_{l_e}^2+v_{l_\mu}^2}}
{\sqrt{y_\tau^2 v_d^2+\lambda_\tau^2(v_{l_e}^2+v_{l_\mu}^2)}}&
\displaystyle\frac{y_\tau v_d}{\sqrt{y_\tau^2 v_d^2+
\lambda_\tau^2(v_{l_e}^2+v_{l_\mu}^2)}} 
\end{pmatrix}.  
\end{equation}

\section{Lepton mixing matrix}

The lepton mixing matrix is $V=U_l^\dag U_\nu$.  It is obtained 
that $\nu_e-\nu_\mu$ mixing is
\begin{equation}
\label{mixing}
V_{e2}=\frac{v_{l_\mu}^2-v_{l_e}^2}{v_{l_e}^2+v_{l_\mu}^2} 
\displaystyle e^{-i\frac{\beta_1}{2}} \,.
\end{equation}  
The $\nu_\mu-\nu_\tau$ mixing is
\begin{small}
\begin{equation}
\label{mixing2}
\begin{split}
V_{\mu3}=& \displaystyle\frac{2v_{l_e}v_{l_\mu}v_{l_\tau}}
{\sqrt{v_{l_e}^2+v_{l_\mu}^2}(v_{l_\tau}^2+\Delta\lambda)}
\displaystyle\frac{y_\tau v_d}
{\sqrt{y_\tau^2 v_d^2+\lambda_\tau^2(v_{l_e}^2+v_{l_\mu}^2)}}
e^{-i\frac{\beta_2}{2}}\\ 
&\displaystyle -\frac{\lambda_\tau\sqrt{v_{l_e}^2+v_{l_\mu}^2}}
{\sqrt{y_\tau^2 v_d^2+\lambda_\tau^2(v_{l_e}^2+v_{l_\mu}^2)}}
e^{-i\delta_{v_d}-i\frac{\beta_2}{2}}\,.
\end{split}
\end{equation}
\end{small}
The $\nu_e-\nu_\tau$ mixing is
\begin{equation}
V_{e3}\simeq\displaystyle\frac{v_{l_\mu}^2-v_{l_e}^2}
{\sqrt{v_{l_e}^2+v_{l_\mu}^2}}
\frac{v_{l_\tau}}{v_{l_\tau}^2+\Delta\lambda}e^{-i\frac{\beta_2}{2}}\,.
\end{equation}  
Experimental data for best values of these mixings are 
$|V_{e2}|\simeq 0.54$, $|V_{\mu3}|\simeq 0.65$, and 
$|V_{e3}|\simeq 0.15$ \cite{pdg}.  Obviously, taking 
$v_{l_\mu}\simeq 2v_{l_e}$, $|V_{e2}|$ is in agreement with data.  The 
value of $v_{l_\tau}$ is taken to be larger and still in the natural 
range, $v_{l_\tau}\simeq 3v_{l_\mu}$.  Choosing 
$\Delta\lambda \simeq 0.3v_{l_\tau}^2$, it is easy to get 
$|V_{e3}|\simeq 0.3|V_{e_2}|$.  

For $|V_{\mu3}|$, there are two terms in Eq.(\ref{mixing2}), neglecting 
the first term for simplicity, this mixing would be maximal if 
$\lambda_\tau\sqrt{{v_{l_\mu}}^2+{v_{l_e}}^2}=y_\tau v_d$, namely 
$\lambda_\tau\simeq 0.8$.  Of course, a smaller $\lambda_\tau$ is more 
natural.  Therefore this model slightly favors the atmospheric neutrino 
angle to be in the first octant. 

The important CP violation in neutrino oscillations is given through 
the invariant parameter $J$ \cite{j}, 
\begin{equation}
\label{jar}
\Im(V_{i\lambda}V_{j\rho}V_{i\rho}^*V_{j\lambda}^*)=
J\sum_{\kappa,\delta}\epsilon_{ijk}\epsilon_{\lambda\rho\delta},
\end{equation}
and
\begin{small}
\begin{equation}
\begin{split}
J \simeq 
&\displaystyle\frac{2v_{l_e}v_{l_\mu}v_{l_\tau}(v_{l_e}^2-v_{l_\mu}^2)^2 
\lambda_\tau y_\tau v_d}{(y_\tau^2 v_d^2+\lambda_\tau^2(v_{l_e}^2
+v_{l_\mu}^2))(v_{l_e}^2+v_{l_\mu}^2)^2(v_{l_\tau}^2+\Delta\lambda)}
\sin\delta \simeq 0.04 \sin\delta,\\[3mm] 
\delta=&-\delta_{v_d} \,.
\end{split}
\end{equation}
\end{small}
$\delta_{v_d}$ is expected to be large, namely $|\sin\delta|\sim 0.1-1$.  
This agrees with current preliminary experimental results 
\cite{t2k-nova}.

\section{Majorana neutrino mass}

The effective Majorana mass in the neutrinoless double beta decay is 
\begin{small}
\begin{equation}
\langle m\rangle_{ee}=|m_{\nu_1}{V_{e1}}^2+m_{\nu_2}{V_{e2}}^2+m_{\nu_3}{V_{e3}}^2|\,.
\end{equation}
\end{small}
In this work, it is 
\begin{equation}
\langle m\rangle_{ee} = 
\left|m_{\nu_1}|V_{e1}|^2 + m_{\nu_2}|V_{e2}|^2 + 
\displaystyle m_{\nu_3}|V_{e3}|^2 e^{i(\delta_{\lambda_1}-\beta_2)}
\right| 
\simeq 0.02 ~{\rm eV}\,.  
\end{equation}
In the above formula, the $V_{e3}$ term has a Majorana phase 
dependence, which is negligibly small anyway.

\section{Discussions}

Like gauge theories which are used to describe the elementary particle 
interactions, SUSY is used for fermion masses.  Our model is the minimal 
SUSY SM with a vector-like triplet field extension, but SUSY breaks at a 
high scale and the R-parity (lepton number) is not required.  The 
sneutrino VEVs result in a neutrino mass which is suppressed by the 
Zino mass.  This is a nice realization of the type-I seesaw mechanism 
which, even does not need to introduce any right-handed neutrino.  The 
triplet field is originally for the realistic Higgs mass.  However, it 
also contributes to neutrino masses through a type-II seesaw mechanism.  
The Zino related seesaw mechanism results in only one massive neutrino. 
By including the triplet contribution, the neutrino masses can be 
realistic.  Compared to our previous studies \cite{2,3},  a more natural 
pattern for neutrino masses is obtained. 

To be numerically natural, let us return back to the original 
superpotential in the beginning.  The couplings are assumed to be taken 
natural values.  The field VEVs are mainly fixed by the soft parameters 
in the Lagrangian, in addition to those in the superpotential.  To fit 
the lepton spectrum and mixing, we take $v_{l_e}\simeq 1$ GeV, 
$v_{l_\mu}\simeq 2$ GeV, $v_{l_\tau}\simeq 6$ GeV, $v_d\simeq 10$ GeV, 
and $v_u\simeq 228$ GeV.  Note $v_{l_\tau}$ does not break the flavor 
symmetry, it is natural that its value is more close to $v_d$.  And the 
large $v_u/v_d$ ratio is for explaining the top quark mass \cite{2}.  
When $\lambda_1^\prime$, $\lambda_2^\prime$ and $v_{l_\tau}^2$ are in 
the same order, the correct neutrino spectrum is obtained.  In terms of 
parameters in the superpotential, we have 
$M_{\tilde{Z}}\simeq 3\times 10^{11}$ GeV.  
$M_T \simeq (1-10) M_{\tilde{Z}}$, and $\lambda$'s $\simeq (0.01-0.1)$.  

It is necessary to check the reliability of our approximation in 
estimating the neutrino masses.  That approximation about the phases 
can be good when the quantities appear in the mass formulae are 
hierarchical, say if $\lambda^\prime_1 \gg v_{l_e}^2+v_{l_\mu}^2$.  As 
it has been seen that this is indeed the case for $m_{\nu_2}$.  In 
$m_{\nu_3}$ (Eq.(\ref{m3})), $\lambda^\prime_2$, $\lambda^\prime_1$ and 
$v_{l_\tau}^2$ are of the same order.  This allows us to look at an 
extreme case where the phase is $\pi$.  In this case, there is a 
possibility of inverted neutrino mass hierarchy, namely a very small 
$m_{\nu_3}$.  But this is achieved through a large cancellation between 
$\lambda^\prime_2$ and $v_{l_\tau}^2$.  Although this is possible, it 
is unnatural.  

The physics of neutrinos in this work is quite different from that in 
Refs. \cite{2,3}.  This is mainly due to the triplet.  In Ref. \cite{2}, 
we introduced a singlet, the neutrino mass matrix $M_1^\nu$ was that 
with only the $33$ matrix element nonvanishing.  And in \cite{3}, the 
triplet replaced the singlet for the Higgs mass in the beginning, 
however, in the neutrino mass analysis, we took $\lambda_1^\nu$ to be 
zero which essentially was the same as that for the singlet case.  
Taking $\lambda_1^\nu$ to be zero was actually unreasonable because our 
principle is to treat all the basic couplings close to $0.01-1$.  As a 
result, in Refs. \cite{2,3}, there was always one massless neutrino.  
That led to that the Majorana mass $\langle m\rangle_{ee}$ is about 
$10^{-3}$ eV.  In addition, in \cite{3} it was wrong to say CP violation 
is small in the lepton sector.

\section{Summary}  

In summary, in the model of high scale SUSY for understanding the 
fermion mass hierarchies, we have studied CP violation in the lepton 
sector, and other aspects of neutrino physics in detail.  In the 
analysis, the phases of the Higgs and sneutrino VEVs, and contribution 
of the $\lambda_1^\nu$ term in superpotential (\ref{2}), have been 
included.  This analysis is more complete than previous consideration.  
The neutrino mass matrix, and the charged lepton one, are fixed by the 
model.  Its specific feature is the triplet contribution, the approximate 
degeneracy of neutrinos $\nu_1$ and $\nu_2$ can be naturally explained.  

This model could not predict exact values of the fermion masses because 
of the flavor symmetry breaking as well as SUSY breaking.  However, the 
principle we follow is that all the coupling constants should be in the 
natural parameter range which is about $(0.01-1)$.  Taking triplet 
contribution dominant, and inputting relevant 
experimental data on leptons, we obtain that 
(i) $m_{\nu_1} \simeq 0.020$ eV, $m_{\nu_2} \simeq 0.022$ eV, 
$m_{\nu_3} \simeq 0.054$ eV.  This normal ordering neutrino spectrum is 
to be checked in JUNO experiment \cite{juno}.     
(ii) CP violation in neutrino oscillation most probably is large.  There 
have been some experimental hint on this \cite{t2k-nova}.  CP violation 
in neutrino oscillations is a great study task experimentally 
\cite{future-cp}.  
(iii) The effective Majorana neutrino mass in the neutrinoless double 
beta decay is about $0.02$ eV, it is within the detection ability of 
future measurements \cite{majorana}.  
(iv) $\theta_{23}$ is slightly favored being in the first octant.    
(v) The electron neutrino mass to be measured in $\beta$ decays is 
about $0.02$ eV.  This is, however, still one order of magnitude lower 
than the future limit of direct measurements \cite{katrin}.  
(vi) The sum of three neutrino masses is close to 
$\sum{m_\nu}\simeq 0.1$ eV.  If the standard cosmology is correct, 
astrophysics measurements on the cosmic microwave background has 
constrained this sum to be $< 0.15$ eV \cite{planck}.  It is interesting 
to note that a recent analysis showed the sum is about $\sim 0.11$ eV 
\cite{sum}.  
Most of the above predictions are close to their experimental limits, 
therefore, this model will soon be checked experimentally.

\begin{acknowledgments}
We would like to thank Gui-Jun Ding and Zhen-hua Zhao for very helpful 
discussions.  This work was supported in part by the National Natural 
Science Foundation of China (No. 11375248 and 11875306).
\end{acknowledgments}

\end{document}